\documentclass[11 pt,a4 paper]{article}
\usepackage{jheppub}
\pdfoutput=1

\usepackage{amsmath}
\usepackage{amssymb}
\usepackage{amsfonts}
\usepackage{braket}
\usepackage[utf8x]{inputenc}

\DeclareMathOperator{\Tr}{Tr}

\title{Multipartite Reflected Entropy}
\author[a,b]{Ning Bao}
\author[a]{and Newton Cheng}

\affiliation[a]{Center for Theoretical Physics and Department of Physics, University of California, \\Berkeley, CA, 94720, USA}
\affiliation[b]{Computational Science Initiative, Brookhaven National Lab, \\Upton, NY, 11973, USA}

\emailAdd{ningbao75@gmail.com}
\emailAdd{newtoncheng@berkeley.edu}

\abstract{We discuss two methods that, through a combination of cyclically gluing copies of a given $n$-party boundary state in AdS/CFT and a canonical purification, creates a bulk geometry that contains a boundary homologous minimal surface with area equal to 2 or 4 times the $n$-party entanglement wedge cross-section, depending on the parity of the party number and choice of method. The areas of the minimal surfaces are each dual to entanglement entropies that we define to be candidates for the $n$-party reflected entropy. In the context of AdS$_3$/CFT$_2$, we provide a boundary interpretation of our construction as a multiboundary wormhole, and conjecture that this interpretation generalizes to higher dimensions.}

\begin{document}
\maketitle

%%-------- Manuscript ------------------------------------------------------
\section{Introduction}
Given a bipartite pure state $\rho_{AB} = \ket{\psi}\bra{\psi}$ on a factorized Hilbert space $\mathcal{H} = \mathcal{H}_A\otimes\mathcal{H}_B$, the von Neumann entropy of one of the reduced density matrices $\rho_A = \Tr_B\rho_{AB}$ quantitatively captures the classical and quantum correlations between $\mathcal{H}_A$ and $\mathcal{H}_B$:
\begin{equation}
	S(A) = -\Tr\rho_{A}\log\rho_{A}.
\end{equation}
This entropy is commonly called the \emph{entanglement entropy}, and it has become a ubiquitous quantity in the study of holography, where the Ryu-Takayanagi (RT) prescription \cite{Ryu:2006bv,Hubeny:2007xt} relates the entanglement entropy of a CFT subregion simply to the area of a particular minimal surface $\mathcal{M}$ in its dual AdS geometry:
\begin{equation}
	S(A) = \frac{\mathcal{A}[\mathcal{M}]}{4G_N}.
\end{equation}
It would not be unreasonable to think that other information-theoretic quantities defined on a boundary CFT could be similarly dual to other bulk quantities. Moreover, the von Neumann entropy fails to be a good measure of ``total'' entanglement for mixed states or states with more than 2 parties, so finding similar holographic relationships for other entanglement measures that are better able to probe such classes of systems would be of particular interest.

One recent example in this direction is the conjecture that the entanglement of purification $E_P$ \cite{Terhal:2002} is computed by the area of a particular bulk minimal surface called the entanglement wedge cross-section $E_W$ \cite{Takayanagi:2017knl, Nguyen:2017yqw}. This $E_P = E_W$ conjecture, as well as other candidates for the information quantity dual to $E_W$, has since been the subject of a large body of work \cite{Bao:2017nhh, Hirai:2018jwy, Espindola:2018ozt, Bao:2018gck, Umemoto:2018jpc, Kudler-Flam:2018qjo, Tamaoka:2018ned, Bao:2018fso, Agon:2018lwq, Caputa:2018xuf, Kudler-Flam:2019oru, Du:2019emy, Jokela:2019ebz, Bao:2019wcf, Harper:2019lff, Kudler-Flam:2019wtv, Kusuki:2019rbk, Kusuki:2019zsp, Umemoto:2019jlz, Jeong:2019xdr, Levin:2019krg, Kusuki:2019evw}, in part for the reasons stated above. 

For generic quantum bipartite mixed states $\rho_{AB}$, computing the entanglement of purification is an optimization problem:
\begin{equation}
	E_P(A:B) = \inf\limits_{\ket{\psi}_{ABA^\star B^\star}}S(AA^\star),
\end{equation}
where the minimization is taken over all purifications $\ket{\psi}$ of $\rho_{AB}$ and corresponding auxiliary Hilbert spaces $A^\star B^\star$. Unfortunately, this means that it is difficult to directly prove whether $E_P$ equals $E_W$, due to the difficulty of the boundary computation. Perhaps motivated by this difficulty, an alternative boundary dual to $E_W$ was developed in \cite{Dutta:2019gen}: the reflected entropy $S_R$, which is a von Neumann entropy related to $E_W$ by
\begin{equation}
	S_R(A:B) = \frac{\mathcal{A}[\mathcal{M}_R]}{4G_N} = 2E_W(A:B),
\end{equation}
where $\mathcal{M}_R$ is the \emph{reflected minimal surface} in an algorithmically-constructed bulk geometry. Because $S_R$ does not involve a minimization over purifications, it is easier to directly compute than $E_P$, e.g. via path integral.

In this paper, we extend the construction in \cite{Dutta:2019gen} to the multipartite case with arbitrarily chosen finite party number $n$. We will provide two topologically-distinct bulk geometries that each contain a minimal surface whose area computes either twice or four times the $n$-party $E_W$. These geometries are constructed in two steps: a ``replica'' step that serves to generate multipartite entanglement by gluing copies of the original bulk geometry together using the gluing constructions discussed in \cite{Engelhardt:2018kcs}, and the doubling step developed in \cite{Dutta:2019gen} to purify the construction. We then argue that the boundary interpretation of these bulk geometries, at least in three dimensions, is as a multiboundary wormhole with low-party number entanglement of the kind described in \cite{Balasubramanian:2014hda}.\footnote{During the completion of this draft we learned of an independent forthcoming work by Jonathan Harper that may provide another generalization of the reflected entropy to multiple parties.}

The paper is organized as follows: in Section \ref{prelim}, we review the multipartite entanglement wedge cross-section and the reflected entropy for bipartite systems. In Section \ref{bulk}, we introduce two procedures to generate pure bulk geometries from a given $n$-party CFT state with an AdS dual, such that the von Neumann entropy of a particular combination of boundary regions is an integer multiple of the $n$-partite $E_W$. In Section \ref{boundary}, we present an argument that, in three dimensions, the boundary dual of our construction is a multiboundary wormhole. We conclude in Section \ref{disc} by discussing some interesting aspects and future directions for our work.

One note worth mentioning at the onset is that throughout this work we will be mainly considering moments of time reflection symmetry for simplicity, though by previous arguments on $E_W$ and the fact that the techniques we use are fully covariant, we expect our results to generalize easily to the covariant setup. The main consequence of working in this limit will be that we will use the terms ``minimal surface'' and ``extremal surface'' interchangeably unless otherwise specified.

During the editorial process, another independent proposal for the multipartite reflected entropy was released \cite{Chu:2019etd}.

\section{Preliminaries}\label{prelim}
\subsection{Entanglement wedge cross-section}
\begin{figure}
	\centering
	\includegraphics[width=0.5\linewidth]{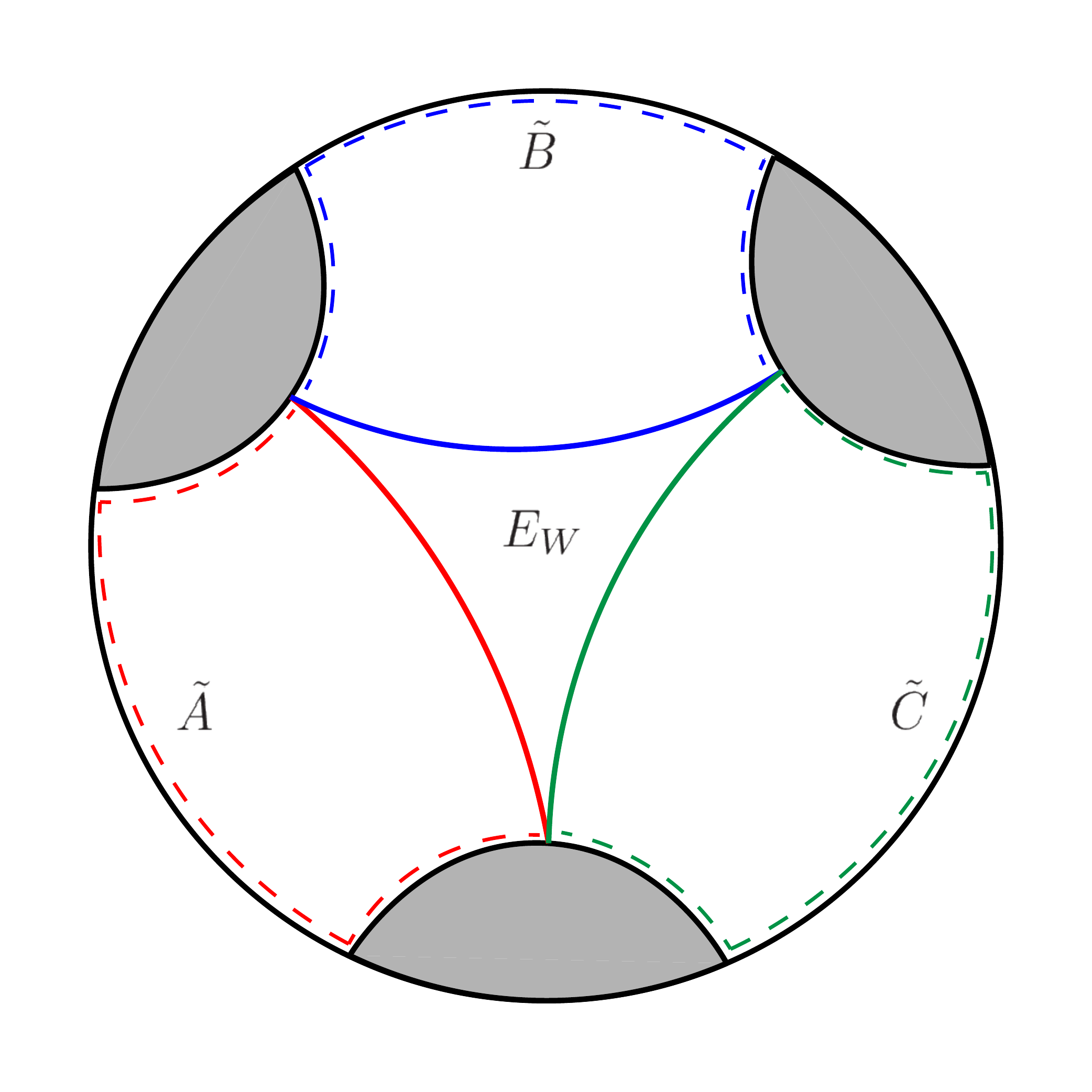}
	\caption{An example of the $E_W$ surface for a 3-party boundary state. The homologous surfaces that form the $E_W(A:B:C)$ surface are shown in solid lines. The 3 corresponding surfaces $\tilde{A},\tilde{B},\tilde{C}$ are shown in dashed lines, containing their respective boundary regions, as well as part of the minimal surfaces that separate the boundary regions. }
	\label{fig:multpartewcs}
\end{figure}

We first briefly review the construction of the $n$-party entanglement wedge cross-section in \cite{Umemoto:2018jpc}. Begin with a CFT divided into $n$ boundary subregions $A_1,\ldots, A_n$. There are $n+1$ minimal surfaces of interest: $\Gamma_{A_1},\ldots,\Gamma_{A_n}, \Gamma_{A_1\ldots A_n}$, whose areas compute $S_{A_1},\ldots,S_{A_n},S_{A_1\ldots A_n}$, respectively. Now divide the boundary into $n$ new regions $\{\tilde{A}_1,\ldots,\tilde{A}_n\}$ whose union is equal to the union of all boundary subregions and $\Gamma_{A_1\ldots A_n}$ (this is also the boundary of the entanglement wedge on the given time-slice):
\begin{equation}
	\tilde{A}_1\cup\ldots\cup \tilde{A}_n = A\cup\ldots\cup A_n \cup \Gamma_{A_1\ldots A_n}
\end{equation}
with the condition that the new regions contain the old ones:
\begin{equation}
	A_i \subseteq \tilde{A}_i,\text{ for all }i,
\end{equation}
and denote the boundary of the new regions
\begin{equation}
	\mathcal{D}_{A_1\ldots A_n} \equiv \partial (\tilde{A}_1\cup\ldots\cup \tilde{A}_n).
\end{equation}
We then choose a set of $n$ surfaces $\{\Sigma_{A_1},\ldots,\Sigma_{A_n}\}$ whose union we denote
\begin{equation}
\Sigma_{A_1\ldots A_n} \equiv\Sigma_{A_1} \cup \ldots \cup \Sigma_{A_n},
\end{equation}
and require that they satisfy:
\begin{enumerate}
	\item $\Sigma_i$ is homologous to $\tilde{A}_i$ inside the entanglement wedge of the full boundary system $A_1\cup\ldots \cup A_n$
	\item $\partial \Sigma_{A_1\ldots A_n} = \mathcal{D}_{A_1\ldots A_n}$
\end{enumerate}
The entanglement wedge cross-section is then defined as the area in Planck units of $ \Sigma_{A_1\ldots A_n}$, minimized over all choices of $\{\tilde{A}_1,\ldots,\tilde{A}_n\}$:
\begin{equation}
	E_W(A_1:\ldots :A_n) = \min\limits_{\tilde{A}_1,\ldots,\tilde{A}_n}\frac{\mathcal{A}[\Sigma_{A_1\ldots A_n}]}{4G_N}.
\end{equation}
An example is shown in figure \ref{fig:multpartewcs}. 
This quantity is conjectured to compute the multipartite entanglement of purification of the boundary subregions \cite{Umemoto:2018jpc,Bao:2018gck}:
\begin{equation}\label{multep}
	E_W(A_1:\ldots :A_n) = \frac{1}{2}\min\limits_{\ket{\psi}_{A_1A_1^\star\ldots A_nA_n^\star}}\sum_{i=1}^nS(A_iA_i^\star) = E_P(A_1:\ldots:A_n),
\end{equation}
where the minimization is over all purifications $\ket{\psi}_{A_1A_1^\star\ldots A_nA_n^\star}$ of the original state $\rho_{A_1\ldots A_n}$.

We note that definition of the multipartite $E_W$ in \cite{Umemoto:2018jpc} reduces to twice the entanglement wedge cross-section defined in \cite{Takayanagi:2017knl}. Unless otherwise stated, we will use the definition in \cite{Takayanagi:2017knl} when $n = 2$, and the definition in \cite{Umemoto:2018jpc} for $n > 2$.

\subsection{Bipartite reflected entropy}
\begin{figure}
	\centering
	\includegraphics[width=\linewidth]{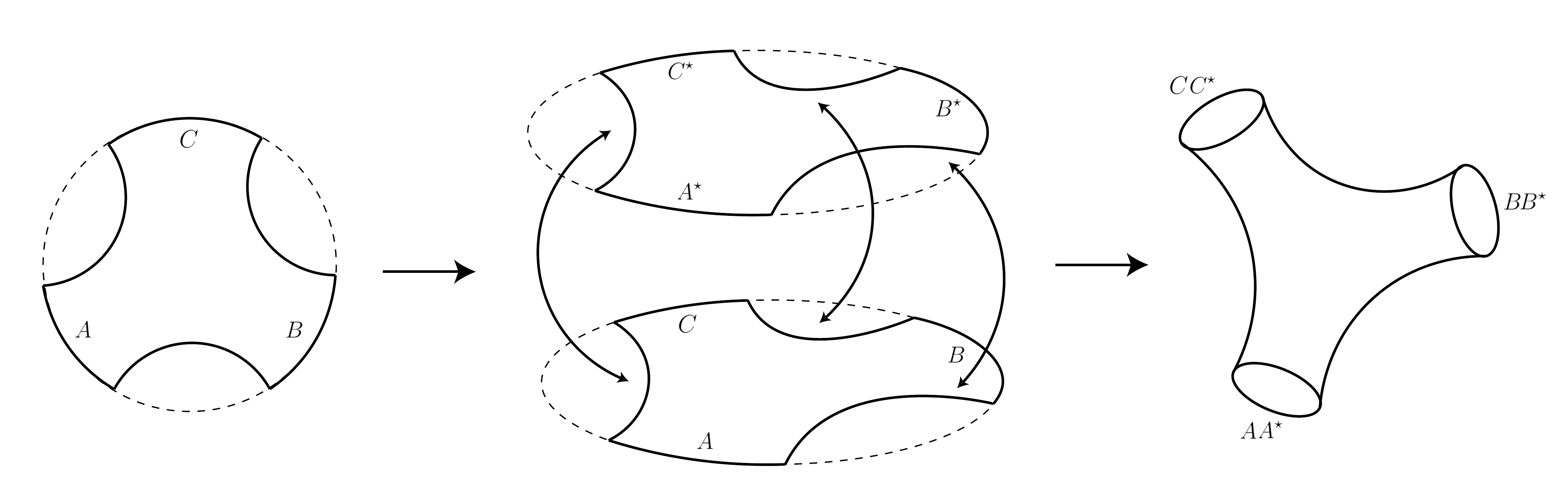}
	\caption{The bulk interpretation of the canonical purification for a 3-party boundary state. The geometry is doubled and glued to the original geometry along identical minimal surfaces that separate the boundary regions. The result is a ``pair of pants'' topology with 3 asymptotic regions. From a boundary perspective, this geometry corresponds to a 3-boundary wormhole in the context of AdS$_3$/CFT$_2$. As we note below, simply doubling the original geometry leads to a reflected entropy defined on the glued geometry will generically fail to capture multipartite correlations.}
	\label{fig:n=3 doubling}
\end{figure}

We now review the bipartite reflected entropy introduced in \cite{Dutta:2019gen}. The basis of the reflected entropy is a canonical purification for a generic bipartite state $\rho_{AB} \in \mathcal{H}_{AB}$. Fixing an appropriate orthonormal basis for $\rho_{AB}$, this purification doubles the Hilbert space:
\begin{equation}
	\rho_{AB} = \sum_ip_i\ket{\psi_i}\bra{\psi_i} \quad \overset{\text{purification}}{\longrightarrow} \quad \ket{\sqrt{\rho_{AB}}} = \sum_i\sqrt{p_i}\ket{\psi_i}_{AB}\ket{\psi_i}^*_{A^\star B^\star}.
\end{equation}
Note that this purification procedure is independent of the state's party number. A simple example is the thermofield double state, which is the canonical purification of the thermal state with inverse temperature $\beta$:
\begin{equation}
	\rho_{\text{thermal}} = \sum_ie^{\beta E_i}\ket{i}\bra{i}  \quad \overset{\text{purification}}{\longrightarrow} \quad \ket{\text{TFD}} =  \sum_ie^{\beta E_i/2}\ket{i}\ket{i}^*
\end{equation}
The reflected entropy is then defined as a von Neumann entropy across $AA^\star$:
\begin{equation}
	S_R(A:B) = S(AA^\star) = -\Tr \rho_{AA^\star}\log \rho_{AA^\star},
\end{equation}
where $\rho_{AA^\star} = \Tr_{BB^\star}\ket{\sqrt{\rho_{AB}}}\bra{\sqrt{\rho_{AB}}}$. For a holographic system, the bulk interpretation of the canonical purification is a CPT doubling of the original bulk geometry, which is then glued to the original geometry along minimal surfaces that separate the boundary regions using the construction developed in \cite{Engelhardt:2018kcs}. An example of this is shown in figure \ref{fig:n=3 doubling}.
The construction essentially states that one is always free to glue CPT-conjugate spacetimes along identical minimal surfaces\footnote{Indeed, any two spacetimes that share an identical extremal surface and have matching twists allow for such a gluing, though in this work we will only need the weaker statement in the main body to prove our result.}, in the sense that the resulting spacetime has a continuous metric, a well-defined causal structure, and solves Einstein's equations with a stress-energy tensor that satisfies the Null Energy Condition. 

One may then apply the RT prescription in the resulting glued spacetime to compute $S_R$ by computing the area of a ``reflected minimal surface'' $\mathcal{M}_R$:
\begin{equation}
	S_R(A:B) = \frac{\mathcal{A}[\mathcal{M}_R]}{4G_N}.
\end{equation}
By construction, $\mathcal{M}_R$ is the union of two copies of the bipartite entanglement wedge cross-section surface, so that the reflected entropy is related to $E_W$ by
\begin{equation}
	S_R(A:B) = 2E_W(A:B).
\end{equation}
In this way, we may study the dynamics of the entanglement wedge cross-section and its conjectured boundary duals by studying a simple von Neumann entropy and its dual minimal surface.

An obvious, but incorrect, generalization of the two-party case, along the lines of the multipartite entanglement of purification (\ref{multep}), would be:
\begin{equation}\label{bipart}
S_R(A_1:\ldots:A_n) \neq \frac{1}{2}\sum_{i=1}^n S(A_iA_i^\star)
\end{equation}
The problem with this definition is that it is merely the sum of bipartite entropies, and therefore does not contain any contributions from multipartite entanglement. Therefore, we should not expect that the sum should be proportional to the entanglement wedge cross-section, except in the special case when the original state is bipartite and the right side of (\ref{bipart}) only has two terms. From the bulk perspective, this is the statement that simply doubling the geometry, as in \cite{Dutta:2019gen}, is insufficient to capture the multipartite entanglement as a reflected entropy. As an example, the bulk doubling procedure shown in figure \ref{fig:n=3 doubling} for an $n=3$ state will produce a reflected entropy that only captures the bipartite correlations of the state. As we will show in the next section, this problem can be rectified by considering more copies of the geometry.\footnote{We thank Jonathan Harper for helpful discussions on this point.}

\section{Multipartite reflected entropy: the bulk}\label{bulk}
We now generalize the work done in \cite{Dutta:2019gen} to the multipartite case $n > 2$. Given an $n$-party boundary state, we will require that our multipartite reflected entropy be a von Neumann entropy $S_R$ with two properties:
\begin{enumerate}
	\item It computes the entropy across an algorithmically-constructed bipartite splitting of a purification of the original state.
	\item It is proportional to an integer multiple of the $n$-partite entanglement wedge cross-section
	\begin{equation}
	S_R(A_1:\ldots:A_n) = I(n)E_W(A_1:\ldots:A_n)
	\end{equation}
	where $I(n)$ is some integer that may depend on $n$. 
\end{enumerate}
More precisely, we will assume the validity of the RT formula and find a minimal surface $\mathcal{M}_R$ in some algorithmically-constructed bulk geometry whose area satisfies:
\begin{equation}
\frac{\mathcal{A}[\mathcal{M}_R]}{4G_N} = I(n)E_W(A_1:\ldots:A_n)
\end{equation}
to leading order.

We will present two different candidates for a multipartite reflected entropy. They differ in terms of the topology of the glued geometry in which the entropies are computed, resulting in corresponding differences in the definition of the entropy. For both constructions, we will make use of the CPT gluing procedure developed in \cite{Engelhardt:2018kcs}. Notationally, we will use $A^{i}_j$ to label boundary regions: lower indices correspond to the boundary region, while upper indices correspond to the copy number. For example, $A_1^3$ is the $A_1$ boundary region on copy 3. For simplicity and concreteness, we will mainly consider three-dimensional bulk spacetimes and single-interval boundary regions labeled in a clockwise fashion; as we note below, our methods generalize in a straightforward manner.

\subsection{Candidate 1}
The geometry for Candidate 1 is obtained by gluing copies of the original spacetime cyclically along its minimal surfaces, which we call the replica step. The procedure for constructing Candidate 1 is as follows:
\begin{enumerate}
	\item Choose one minimal surface in the original geometry and glue a CPT copy of the geometry along it
	\item On the copy of the geometry, choose an adjacent\footnote{It is important to stress here that ``adjacency'' is an artifact of three-dimensional bulk spacetimes with each $A_i$ being given by single intervals. We do this to fix a concrete ordering of the regions. In higher dimensions or with boundary regions formed of disjoint unions, this notation can be generalized to a fixed ordering of the disconnected minimal surfaces homologous to the original entanglement wedge, just as one does to define $E_W$ in such situations. Thus, the procedure will continue to work these cases due primarily to the arbitrariness of the ordering.} minimal surface to the glued one and glue another CPT copy of the geometry -- this is critical to ensure that the appropriate minimal surface at the end of the day computes a multiple of $E_W$
	\item Repeat until a loop is completed by gluing the final copy to the original geometry
\end{enumerate}
Naively, this requires $n$ copies for an $n$-partite state, as there are generically $n$ unique (in terms of a fixed labeling scheme) minimal surfaces in such a state, and each one will be glued along once.\footnote{In the case where this is not true, e.g. a tripartite pure state or subregions that are not disjoint, our construction still goes through in exactly the same way. The gluing at the boundary is trivial in the sense that one will just have copies of the spacetime that are not connected. However, we can naturally still define a von Neumann entropy of the boundary regions and purify the construction. The reflected surface will then be the union of disjoint $E_W$ segments on the various copies. We thank Don Marolf for pointing this case out to us.} However, if $n$ is odd, then we cannot close the loop, due to the fact that the $n$th copy of the geometry will not be CPT-conjugate to the first copy. This issue can be dealt with by simply doubling the size of the chain to $2n$ copies in total, after which the gluing procedure goes through the same way, except every unique minimal surface will be glued along twice, albeit in different copies. This procedure then produces a ring-like geometry after the final identification. It is important to note that this glued geometry will generally not be a pure state on the boundary dual -- this is clear from the fact that there will generically still be ``open'' minimal surfaces that have not been glued. To purify the construction, we simply apply the canonical purification, whose bulk dual we know from \cite{Dutta:2019gen} is simply a doubling the space and gluing along those ``open'' minimal surfaces. Notationally, boundary regions in the doubled space will be denoted with a $^\star$.

\begin{figure}
	\centering
	\includegraphics[width=\linewidth]{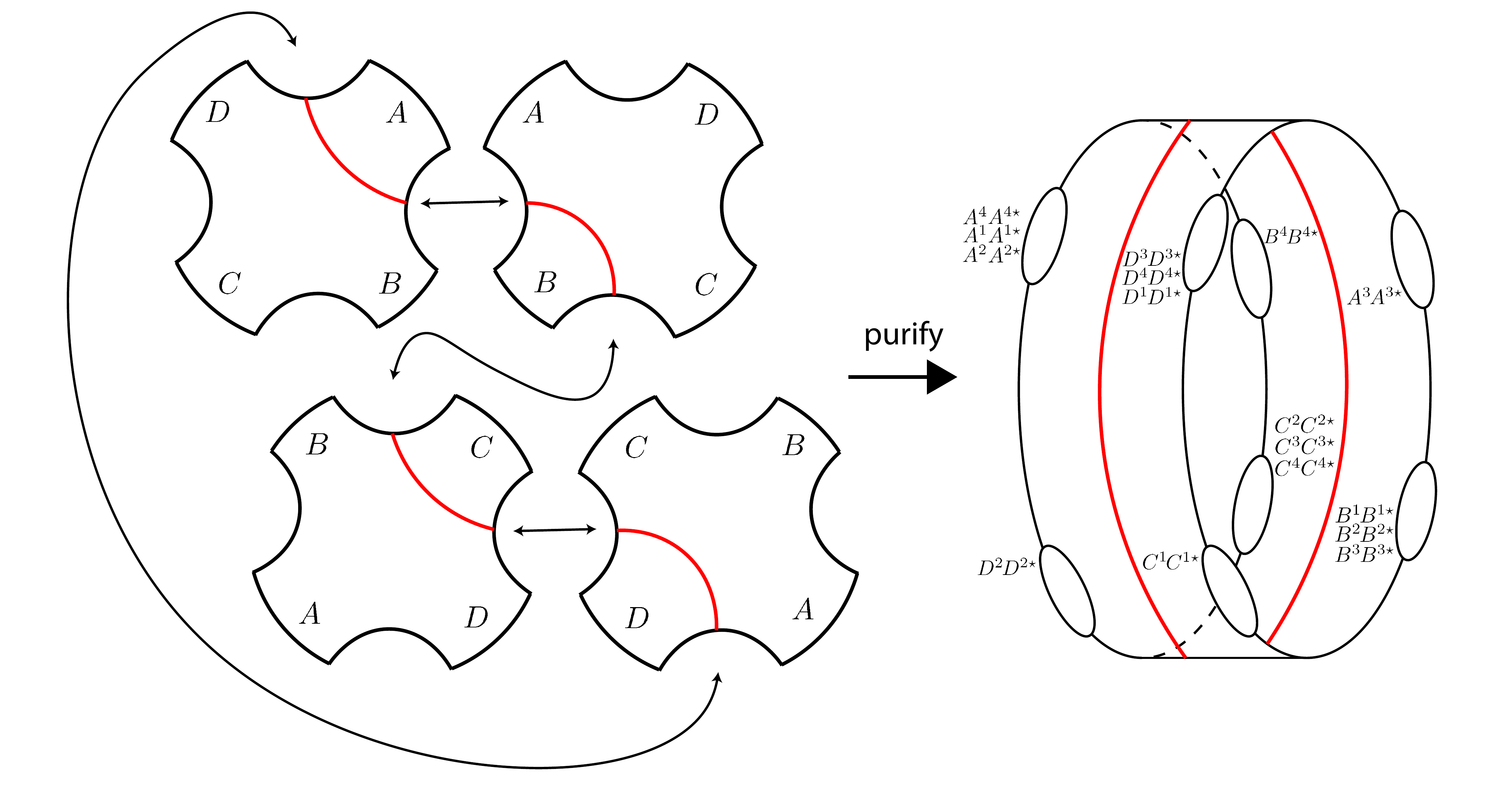}
	\caption{An example of the gluing construction for $n=4$ with the assumptions made in the text. \textit{Left:} The arrows indicate the surfaces that are identified in the gluing. The red lines inside the copies indicate the multipartite $E_W$ surfaces for each boundary region. The sum of the areas in Planck units of these surfaces gives $E_W(A:B:C:D)$. \textit{Right:} After the gluing and the canonical purification, we have a ring that is punctured in various locations. The reflected minimal surface consists of two disconnected pieces, one on the outside of the ring and one on the inside, that each have area $E_W(A:B:C:D)$.}
	\label{fig:n4-gluing}
\end{figure}

Begin the gluing between $A^1$ and $A^2$ along the minimal surface between $A_1$ and $A_2$, and then proceed by gluing $A^2$ and $A^3$ along $A_2$ and $A_3$, and so on. These choices simply correspond to a choice of labels: any gluing that follows the steps above can clearly be obtained by some permutation of boundary and replica labels of any other gluing. We then define the multipartite reflected entropy as the entropy:
\begin{equation}
S_R(A_1:\ldots:A_n) = S(A(n)A^\star(n))
\end{equation}
where we define\footnote{This definition of $A(n)$ is naturally dependent on the choice of label scheme, changing appropriately according to permutations of the labels. Correspondingly, this definition of $A(n)$ is specialized to three dimensions and single-interval boundary regions. The generalization is, again, obtained by replacing the adjacency condition with a more general ordering of the minimal surfaces homologous to the entanglement wedge.}
\begin{equation}\label{A-c1}
A(n) = \begin{cases}
\sum\limits_{\substack{j\text{ odd} \\ i = j\text{ mod } n}}\limits^{n}(A_i^{j-1}A_i^j A_i^{j+1}) \quad  + \sum\limits_{\substack{j\text{ even} \\ i\neq \{j, j+1,j-1\}\text{ mod } n}  }\limits^{n} (A^j_{i}), &n \text{ even} \\
\sum\limits_{\substack{j\text{ odd}\\ i = j\text{ mod } n}}\limits^{2n}(A_i^{j-1}A_i^j A_i^{j+1}) \quad + \sum\limits_{\substack{j\text{ even} \\ i\neq \{j, j+1,j-1\}\text{ mod } n}  }\limits^{2n} (A^j_{i}), &n \text{ odd}
\end{cases}
\end{equation}
The parentheses indicate the groupings of boundaries that combine under the identifications; these are ``half-holes'' in the final geometry that become one of the $n(n-2)$ holes in the final geometry for even $n > 2$, or $2n(n-2)$ for odd $n$. $A(n)$ will have $3n/2 + \frac{n(n-3)}{2} = \frac{n^2}{2}$ terms\footnote{In this context, ``terms'' refers to the number of boundary labels.} if $n$ is even, and $n^2$ terms if $n$ is odd. Naturally, the purity of the state means that we can always use the complement of the definition above to get the same result. An example for $n = 4$ is shown in figure \ref{fig:n4-gluing}, where we have used the boundary labels $A_1 = A$, $A_2 = B$, $A_3 = C$, and $A_4 = D$. In this case, we have:
\begin{equation}
A(n = 4) = (A_1^4A_1^1A_1^2)(A^2_3A_3^3A_3^4)(A^4_2) (A^2_4)= (A^4A^1A^2)(C^2C^3C^4)(B^4)(D^2)
\end{equation}
and the reflected entropy is
\begin{equation}
S_R(A:B:C:D) = S[(A^4A^1A^2A^{4\star}A^{1\star}A^{2\star})(C^2C^3C^4C^{2\star}C^{3\star}C^{4\star})(B^4B^{4\star})(D^2D^{2\star})].
\end{equation}

By construction, the minimal surface $\mathcal{M}_R$ that computes $S_R$ is really the union of the multipartite $E_W$ surfaces that are separately associated to each boundary region. It is worth being very clear about this point: the minimality condition on $\mathcal{M}_R$ is equivalent to the minimality condition on $E_W$. This is immediate from the fact that $\mathcal{M}_R$ is the union of 2 (or 4) identical surfaces, each of which is constructed identically to $E_W$ -- step 2 in the gluing procedure ensures that this is the case. Our construction essentially ``unrolls'' the $E_W$ surface so that there is one segment on one copy of the geometry. These are drawn in figure \ref{fig:n4-gluing}. The individual bipartite $E_W$ surfaces will generally fail to connect to form a closed surface in this procedure for $n\geq2$, and hence will not be boundary homologous to any region in this construction, thus failing to retain an entropic interpretation. It is clear then that the definition of reflected entropy above gives
\begin{equation}\label{reflent}
S_R(A_1:\ldots:A_n) = \frac{\mathcal{A}[\mathcal{M}_R]}{4G_N} = \begin{cases}
2E_W(A_1:\ldots:A_n),&  n\text{ even} \\
4E_W(A_1:\ldots:A_n),&  n\text{ odd}
\end{cases}
\end{equation}
This also demonstrates why we choose to double the copies for odd $n$, rather than work with, say, $n + 1$ copies to deal with the CPT issue -- it would be impossible to get an integer multiple of $E_W$.

Although our result reduces appropriately for the case with $n = 2$, we want to point out that this case is special, because one of either the canonical purification or the gluing step is unnecessary as they become degenerate with each other. This stems from the fact that the $n = 2$ case does not have any multipartite entanglement beyond bipartite entanglement. Geometrically, the gluing procedure will leave one with a pure state, so the canonical purification step does nothing or vice versa. Alternatively, one can interpret this as saying that the gluing procedure is unnecessary, because there is no higher-partite entanglement to be generated.

We have considered a vacuum state for simplicity, but we can easily include black holes -- they simply appear as a modification to the homology constraint on $E_W$, and our construction follows through, in the sense that the final construction will still include a minimal, bipartitioning surface that computes a von Neumann entropy, which we identify as the reflected minimal surface. In this case, there may be more $E_W$ segments than copies of the geometry, because $E_W$ can consist of multiple disjoint surfaces. Rather than each copy contributing one $E_W$ segment to the reflected minimal surface, as with a vacuum state, each copy will contribute multiple $E_W$ segments. The same point holds true for the construction we give in the next part.

\subsection{Candidate 2}
We now present another candidate for the multipartite reflected entropy. In this construction, we go through the same gluing procedure as before, but now we choose one surface to remain unidentified. For concreteness, assume the same labeling and gluing procedure in the previous candidate, and choose the unidentified surfaces to be the ones that would connect $A^{n}$ and $A^1$ along $A_n$ and $A_1$. Without this final identification, the topology after the canonical purification does not resemble a ring, as in the previous case. The minimal reflected entropy in this case is still:
\begin{equation}
S_R(A_1:\ldots:A_n) = S(A(n)A^\star(n)),
\end{equation}
but we now define\footnote{Again, this definition is specific to three dimensions and single-boundary intervals.}
\begin{equation}\label{A-c2}
A(n) = \begin{cases}
(A_1^1A_1^2) + (A_n^n) + \sum\limits_{j > 1,\text{ odd}}\limits^{n-1}(A_j^{j-1}A_j^j A_j^{j+1})\quad  + \sum\limits_{\underset{i\neq \{j, j+1,j-1\}}{j\text{ even,}}  }\limits^{n} (A^j_{i}), &n \text{ even} \\
(A_1^1A_1^2) + (A_n^{n-1}A_n^n) + \sum\limits^{n-2}\limits_{j > 1,\text{ odd}}(A_j^{j-1}A_j^j A_j^{j+1}) \quad  + \sum\limits^{n-1}\limits_{\underset{i\neq \{j, j+1,j-1\}}{j\text{ even,}}  } (A^j_{i}), &n \text{ odd}
\end{cases}
\end{equation}
After purification with odd $n$, the final object has $\frac{1}{2}(n^2 - 3n + 4)$ holes on one side of the reflected minimal surface, and $\frac{1}{2}n(n-1)$ on the other side, for a total of $n^2 - 2n +2$ holes. With even $n$, each side of the reflected minimal surface has $\frac{1}{2}(n^2 - 2n +2)$ many holes. 

In the case that $n$ is even, the combination of boundaries coincides with the first construction, although the topology is different, as evidenced by the fact that the grouping of the boundaries that are glued together is different. The two constructions differ when $n$ is odd -- this comes as no surprise, since the number of copies is different, and that $n$ being odd implies that the boundaries are asymmetric across the minimal reflected surface. For even $n$, there will be $\frac{n^2}{2}$ terms, while for odd $n$, there will be $\frac{n(n-1)}{2}+1$ terms. Due to the purity of the state, we can naturally also use the complement of the definition above to get a definition with $\frac{n(n+1)}{2}-1$ terms.

\begin{figure}
	\centering
	\includegraphics[width=\linewidth]{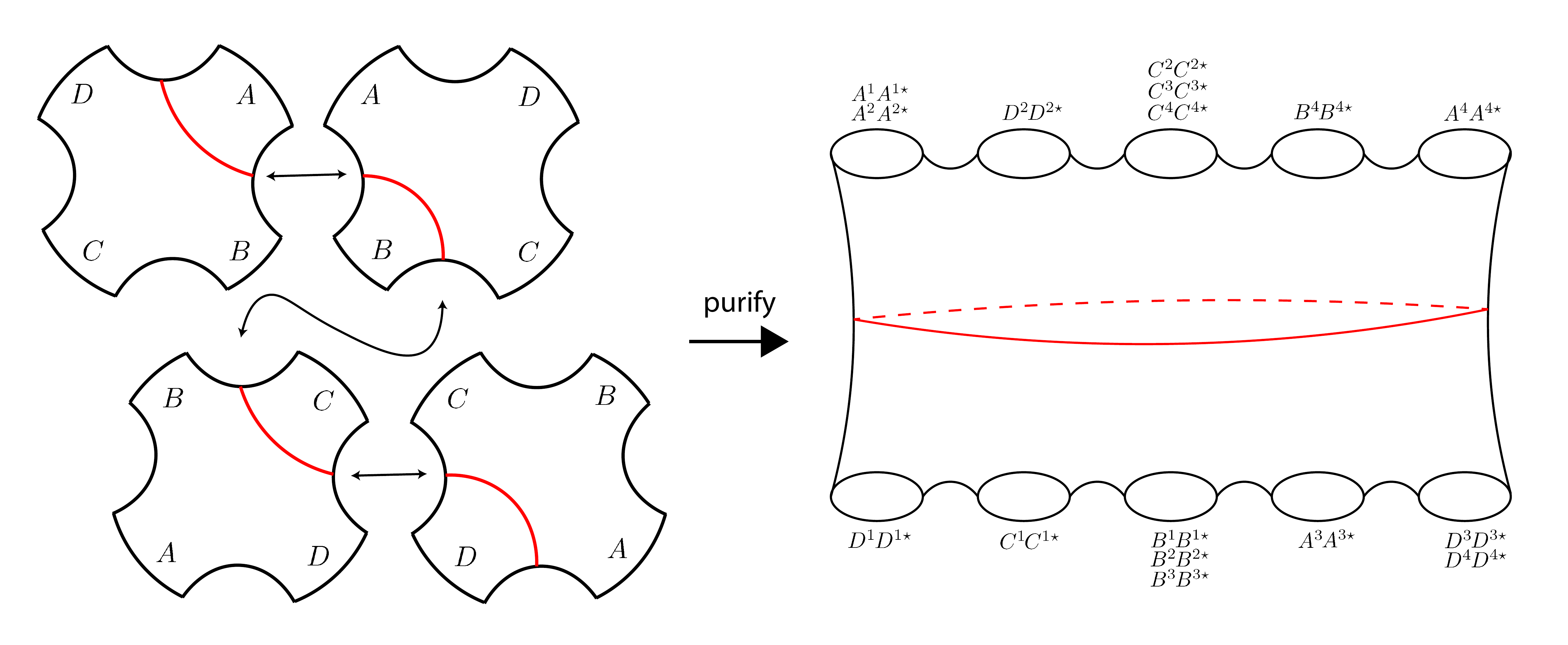}
	\caption{An example of the second candidate for the multipartite reflected entropy for $n=4$. The left figure shows the construction of the glued geometry, and the right figure is the final purified geometry. The $E_W$ surfaces are shown in red. The final glued, purified geometry no longer has the ring-like shape from the first candidate. The reflected minimal surface is one connected piece with area $2E_W(A:B:C:D)$. On comparison to figure \ref{fig:n4-gluing}, it is also clear that this construction has broken the cyclical symmetry of the boundary regions.}
	\label{fig:n4-gluing-2}
\end{figure}

An example with $n=4$ is shown in figure \ref{fig:n4-gluing-2}.
Changing back to the notation $A_1 = A$, $A_2 = B$, $A_3 = C$, and $A_4 = D$, the reflected entropy is given by
\begin{equation}
S_R(A:B:C:D) = S[(A^1A^2A^{1\star}A^{2\star})(A^4A^{4\star})(C^2C^3C^4C^{2\star}C^{3\star}C^{4\star})(D^2D^{2\star})(B^4B^{4\star})].
\end{equation}

By construction, the reflected minimal surface $\mathcal{M}_R$ is a surface whose area can be related to $E_W$ by
\begin{equation}\label{reflent2}
S_R(A_1:\ldots:A_n) = \frac{\mathcal{A}[\mathcal{M}_R]}{4G_N} = \begin{cases}
2E_W(A_1:\ldots:A_n), &n > 2 \\
4E_W(A_1:\ldots:A_n), &n = 2
\end{cases}
\end{equation}
The apparently strange behavior for $n=2$ is a result of our definition of $E_W$ for $n=2$. As we noted previously, the definition of the multipartite entanglement wedge cross-section given in \cite{Umemoto:2018jpc} reduces to twice the original definition of the bipartite entanglement wedge cross-section given in \cite{Takayanagi:2017knl}, so $n=2$ is not a special case if we extend the definition of $E_W$ we use for $n > 2$.

\begin{figure}
	\centering
	\includegraphics[width=\linewidth]{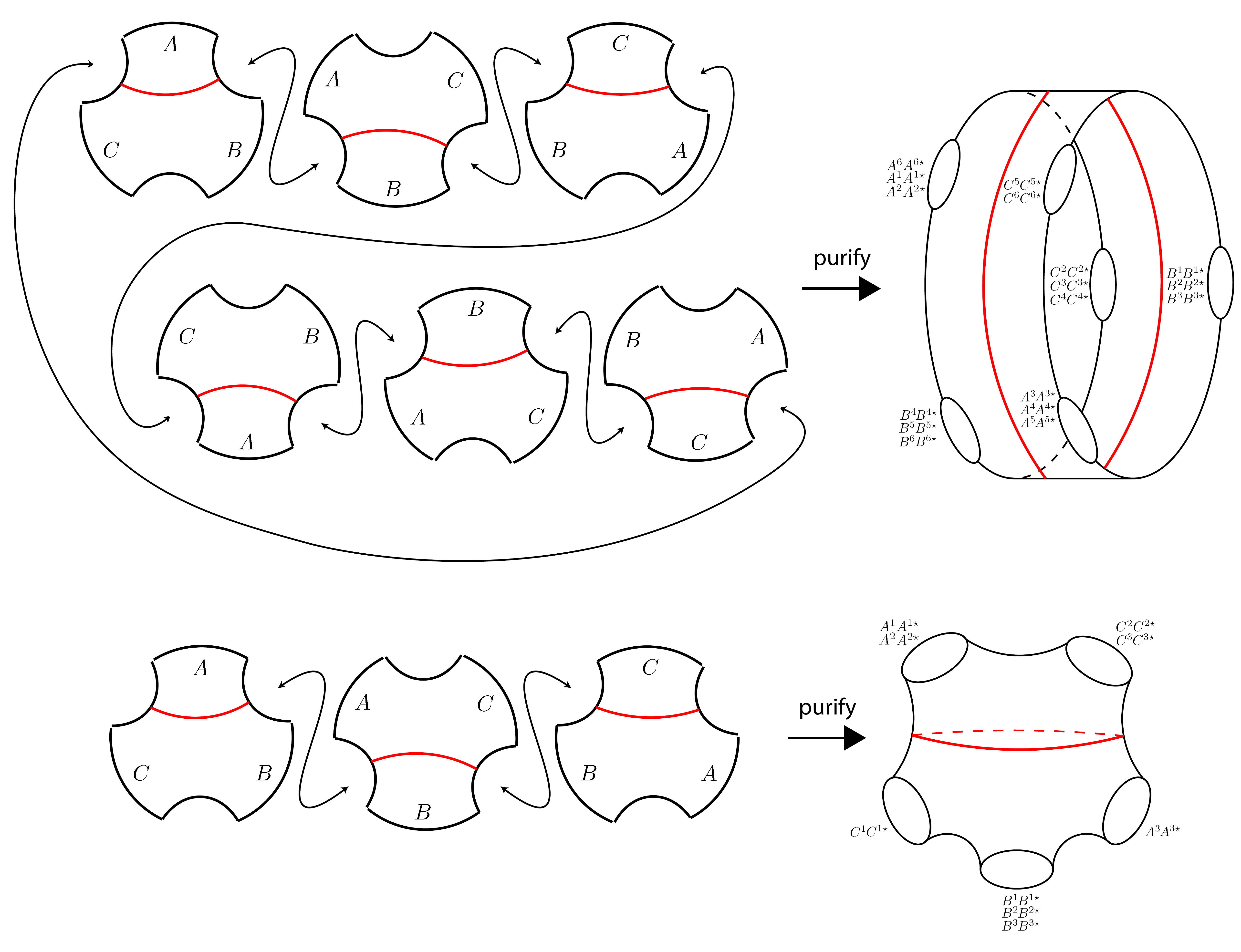}
	\caption{Comparing the two constructions for $n=3$. \textit{Above:} For Candidate 1, we need 6 copies for the replica step, which is then mirrored for a total of 12. The topology is ring-like, and the reflected minimal surface consists of 2 disjoint surfaces, each of which has area equal to $2E_W(A:B:C)$. The reflected entropy is $S_R(A:B:C) = S[(A^6A^1A^2A^{6\star}A^{1\star}A^{2\star})(B^4B^5B^6B^{4\star}B^{5\star}B^{6\star})(C^2C^3C^4C^{2\star}C^{3\star}C^{4\star})]$. \textit{Below:}  For Candidate 2, we only need 3 copies to do the replica step, which is then mirrored for a total of 6. The red lines denote the $E_W$ surfaces on each surface, which combine to form the reflected minimal surface with area $2E_W(A:B:C)$. The reflected entropy is $S_R(A:B:C) = S[(A^1A^2A^{1\star}A^{2\star})(C^2C^3C^{2\star}C^{3\star})]$. }
	\label{fig:n3-gluing}
\end{figure}

Now let us comment on the differences between our two candidates for the multipartite reflected entropy. Candidate 2 has the advantage that, up to a reasonable caveat for $n=2$, obeys $S_R = 2E_W$ for all $n$. Moreover, ``closing" the $E_W$ surface via the canonical purification means the reflected minimal surface is a single surface, as opposed to the two disjoint surfaces in Candidate 1. However, the replica step in Candidate 2 produces a geometry with less symmetry; in particular, it breaks the cyclic symmetry of Candidate 1, because we must choose which surface to leave unidentified. That is, we must choose where to cut the ring produced in the replica step of Candidate 1 (however, we can certainly still permute labels to get a desired labeling scheme). This is clear on comparing the respective forms of $A(n)$ in (\ref{A-c1}) and (\ref{A-c2}). As the simplest example, we can look at the $n=3$ case presented in figure \ref{fig:n3-gluing}. The salient features are the doubled copies in Candidate 1, the decreased symmetry in Candidate 2, and the difference in the topology of the two constructions.

We note here that the equalities in (\ref{reflent}) and (\ref{reflent2}) are leading order statements; they are only expected to be true at $O(N^2)$, with subleading quantum corrections. Similar to the discussion in \cite{Dutta:2019gen}, we expect that the leading order correction arises from bulk entanglement across the reflected minimal surface $\mathcal{M}$:
\begin{equation}
    S_R(A_1:\ldots :A_n) = 2E_W(A_1:\ldots:A_n) + S^{\text{bulk}}(aa^*) + O(1/N^2),
\end{equation}
where $aa^*$ is the bulk region on one side of the reflected minimal surface. To get the correct entropy to all orders, we would need to use the area of the quantum extremal surface analogue of $\mathcal{M}_R$ \cite{Faulkner:2013ana,Engelhardt:2014gca,Dong:2017xht,Engelhardt:2019hmr}, corresponding to the quantum-corrected location of $\mathcal{M}_R$.

\section{Multipartite reflected entropy: the boundary}\label{boundary}
In this section, we provide a qualitative argument that the bulk object giving the reflected entropy is a multiboundary wormhole of the form described in \cite{Balasubramanian:2014hda} for AdS$_3$/CFT$_2$; indeed, the figures we have used are quite suggestive of this description.

The core of our argument is a method for constructing multiboundary wormholes in three dimensions called \textit{doubling} \cite{Brill:1995jv,Brill:1998pr}. Given a vacuum AdS$_3$ timeslice with $n$ boundary subregions separated by boundary-anchored minimal surfaces, one can generate a multiboundary wormhole with $n$ boundaries by taking another copy of the geometry, cutting along the minimal surfaces, and then gluing the remaining geometry. This is precisely what is demonstrated in figure \ref{fig:n=3 doubling}. These wormholes are entirely defined by their boundary regions and the size of their throats, which show up as the areas of particular minimal surfaces in the original geometry. This procedure may seem familiar: it is precisely the bulk construction given in \cite{Dutta:2019gen} as the dual of the canonical purification! In other words, holographic multiboundary wormholes in three dimensions have a boundary interpretation as the canonical purification of a geometry with the appropriate features, i.e. areas of particular minimal surfaces. By virtue of the gluing procedure we use, the unpurified geometry is simply an extended AdS$_3$ vacuum with $n$ many asymptotic boundaries that we double and close to form an $n$-boundary wormhole geometry. 

We note that the wormholes we construct are explicit examples of ``low-partite'' wormholes: the number of boundaries is always greater than the party number $n$ for $n>2$. Indeed, the number of boundaries will be significantly larger than the party number for large $n$ because it scales as $n^2$. As an example, if we consider the $n=3$ case, the final wormhole geometry has 5 or 6 boundaries, depending on the choice of bulk construction, despite there being at most 3-party entanglement for any given choice of boundaries. This agrees with the point made in \cite{Balasubramanian:2014hda} that an $n$-boundary wormhole does not require any intrinsic $n$-party entanglement to have a connected, smooth geometry in the bulk.

Although the doubling argument is technically restricted to vacuum AdS$_3$, the presence of black holes does not change the argument. The horizons simply act as more minimal surfaces that we glue along during the canonical purification, which then become wormholes in the final geometry. This is precisely the case for the thermofield double state, where the canonical purification takes a one-sided black hole to a wormhole geometry.

As noted in \cite{Balasubramanian:2014hda}, the specific boundary state is generically very difficult to write down. Given a generic holographic state on the boundary $\rho$ and its copy dual to a CPT-reflected spacetime $\rho^*$, the boundary interpretation of gluing the two spacetimes along a minimal surface is similarly difficult to explicitly write down. Below, we discuss two alternative methods for determining an explicit form of the boundary state, though they each have their own difficulties. 

\begin{figure}
	\centering
	\includegraphics[width=\linewidth]{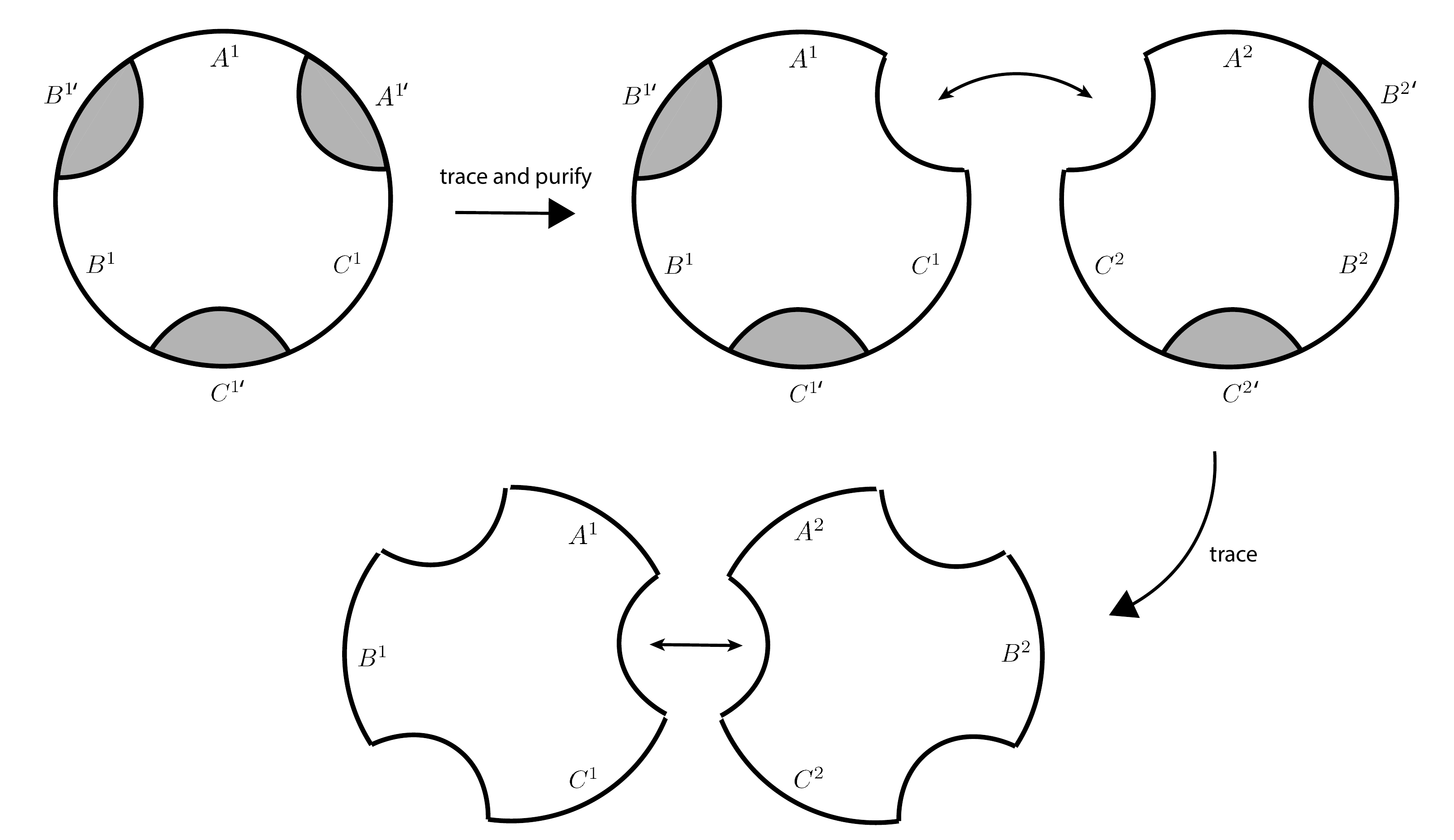}
	\caption{How to compute the boundary state for the gluing of two CPT-conjugate spacetimes using an arbitrary purification on the full boundary Hilbert space. Starting from the pure state in the upper-left, we trace out the subregion ${A^1}'$ of the purifying subsystem that corresponds to the minimal surface we wish to glue. The gluing is done as a canonical purification of the reduced state, as shown in the upper-right. In the last step, the rest of the purifying subsystem is traced out, leaving the glued spacetime.}
	\label{fig:pure_boundary}
\end{figure}

One method to generate the boundary state is to use an assumed purification $\ket{\psi}$ of the initial state $\rho$.\footnote{While in discussion with the anonymous referee, we learned of an independent multipartite reflected entropy proposal \cite{Chu:2019etd} that also uses this method to describe the boundary state.} Given, say, a tripartite state $\rho_{ABC}$, we can always obtain the state by tracing over a pure state $\ket{\psi}_{ABCA'B'C'}$ in the full boundary CFT Hilbert space. Given this state, our gluing procedure is a series of traces and canonical purifications: trace out the region that corresponds to the minimal surface to be glued, perform a canonical purification of the state, then trace out the remainder of the purifying subsystem. This is shown schematically in figure \ref{fig:pure_boundary}. 
In this example we begin with a purification $\ket{\psi}_{A^1 B^1 C^1 {A^1}' {B^1}' {C^1}'}$ and the final state is:
\begin{equation}
    \rho_{A^1 B^1 C^1 A^2 B^2 C^2} = \Tr_{{B^1}' {C^1}' {B^2}' {C^2}'}\ket{\sqrt{\rho_{A^1 B^1 C^1 {B^1}' {C^1}'}}}\bra{\sqrt{\rho_{A^1 B^1 C^1 {B^1}' {C^1}'}}},
\end{equation}
where $\rho_{A^1 B^1 C^1 {B^1}' {C^1}'} = \Tr_{{A^1}'}\ket{\psi}\bra{\psi}_{A^1 B^1 C^1 {A^1}' {B^1}' {C^1}'}$ and $\ket{\sqrt{\rho_{A^1 B^1 C^1 {B^1}' {C^1}'}}}$ is its canonical purification. The procedure can then be recursed as necessary to generate the boundary state of our glued bulk constructions. The difficulty of this method is a requisite knowledge of a specific purification $\ket{\psi}$.

Another method to obtain an explicit boundary state makes use of a recent result regarding the boundary dual to gluing spacetimes whose boundary states are eigenstates of the area operator \cite{Marolf:2019zoo}. The glued state is obtained by ``sewing" the states along the desired minimal surface. If the decomposition of the original state $\rho$ into the basis of fixed-area eigenstates is known, then one can just apply the sewing operation as needed to sew the full states together and find the final boundary state. The difficulty of this method is determining the necessary decomposition for a given state. However, in principle, this only requires knowledge of $\rho$, rather than a purification $\ket{\psi}$.

\section{Discussion}\label{disc}
In this section, we will comment on some other applications and future directions related to the construction of multipartite reflected entropy.

\subsection{$E_W$ winding}
\begin{figure}
		\centering
		\includegraphics[width=0.5\linewidth]{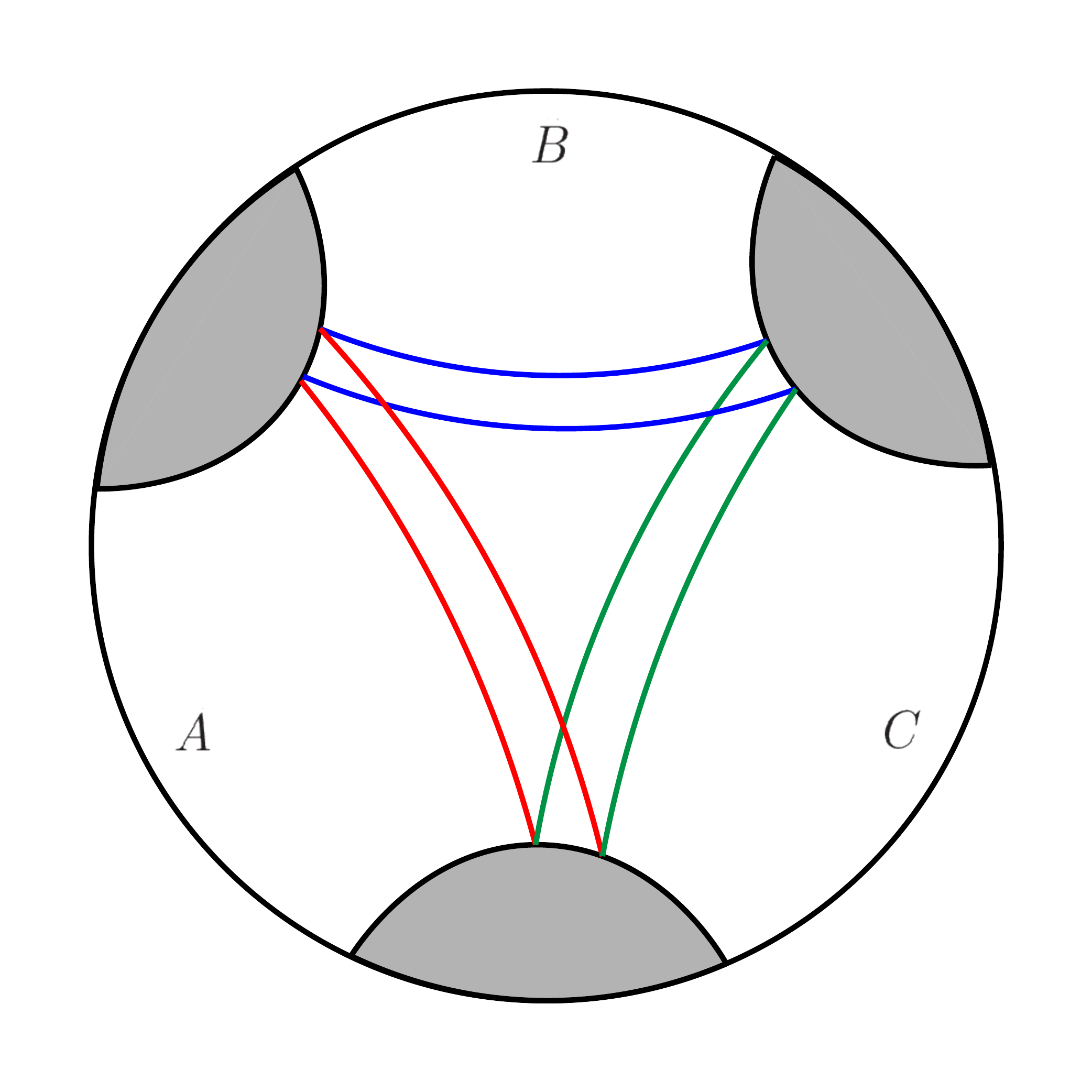}
		\caption{An example of an $E_W$-like surface with winding number 2, so that the surface is composed of 6 segments. We can compute an analogous $S_R$ if we use 6 copies of the geometry, such that each copy contributes one of the segments to the reflected minimal surface.}
		\label{fig:winding}
	\end{figure}

Our construction is able to account for a ``winding number'' parameter in the purification. As discussed in \cite{Harper:2019lff}, there may be exotic cases, usually involving very high curvature, where the $E_W$ surface winds more than once through the geometry, which we show in figure \ref{fig:winding}.
The authors prove that such configurations do not occur in three dimensions\footnote{While this proof is correct, it is possible that such geometries are simply geometric projections of multiple identical geometries on top of each other so that there are intersection points that do not exist in the true, unprojected geometry. As an example, consider a minimal surface in the shape of a cylindrical helix, and then consider a projection in the height coordinate of the cylinder.}, but we just point out that our bulk construction, which works in any finite dimension, still applies and can still compute an appropriate $S_R$. Indeed, our construction can always explicitly build a geometry with a given winding number parameter. Such configurations may be desirable if, for example, one wants to perform a constrained optimization over $E_W$ surfaces with some fixed winding. This is done by simply gluing more copies of the original geometry to take into account the extra segments of the $E_W$ surface, so that when we ``unroll'' the surface, there is still one surface per copy of the geometry. The number of copies needed then just increases by a factor of the winding number. The precise definition of $A(n)$ will change slightly depending on the winding number, but it is straightforward to take into account. 

\subsection{Higher dimensions}
For the sake of being concrete, we have mostly worked in the context of AdS$_3$/CFT$_2$ throughout our analysis, with instances of potential subtlety in higher dimensions pointed out in footnotes. The definition of the multipartite $E_W$ in \cite{Umemoto:2018jpc} and the gluing construction in \cite{Engelhardt:2018kcs} work in arbitrary finite dimensions, so our bulk construction and corresponding analysis for the multipartite reflected entropy should also work in arbitrary finite dimension. Therefore, our candidate definitions for the multipartite reflected entropy remain well-defined if we generalize beyond AdS$_3$/CFT$_2$. Unfortunately, the same cannot be said for our boundary interpretation of the glued, purified geometry. The argument in our analysis of the boundary is restricted to three dimensions because the quotient/doubling procedure for generating multiboundary wormholes is only rigorously understood in three dimensions, and it is unknown if there are similar constructions in higher dimensions. Provided such constructions exist, we expect that a generalization for the argument should hold in higher dimensions, with an appropriate higher-dimensional analogue of the doubling procedure. This would be potentially quite interesting, as it would motivate a two parameter family of purifications of a boundary density matrix dual to an entanglement wedge in arbitrary numbers of dimensions, and indeed motivates the search thereof.

\subsection{Dynamics of $E_W$ and $E_P$}
The main advantage of $S_R$ is that a von Neumann entropy and its corresponding minimal surface are much simpler to study than the entanglement wedge cross-section or the entanglement of purification directly. \cite{Kusuki:2019rbk, Kusuki:2019evw} are examples of recent work in this direction, where the behavior of the bipartite $E_W$ under a local operator quench can be studied with relative ease by directly evaluating the reflected entropy. We expect that the multipartite reflected entropy developed in this paper can serve a similar purpose. Indeed, studying multipartite $E_W$ and $E_P$ is made even more difficult by the fact that generic multipartite entanglement is currently poorly understood, but $S_R$ essentially turns everything into a bipartite correlation. This may allow for a deeper understanding of the entanglement structure of holographic CFT states.

It may even be possible that the simple relationship between $E_P$ and $S_R$ (assuming that $E_P=E_W$ is correct) persists for some class of states away from the holographic limit. In this case, $S_R$ could provide a quite tractable way of computing $E_P$ without ever needing to do the direct and difficult optimization step, by considering these relatively simple to prepare purifications. This could prove to be a direction for fruitful future work.

\acknowledgments
We thank Aidan Chatwin-Davies, Netta Engelhardt, Tom Faulkner, Jonathan Harper, Matt Headrick, Don Marolf, Jason Pollack, and Koji Umemoto for useful comments and interesting discussions. We are grateful to Jonathan Harper for helpful suggestions on a draft of this paper. We also thank the anonymous referee for providing suggestions on where our discussion could be improved. N.C. thanks the Yukawa Institute for Theoretical Physics at Kyoto University for hospitality during the workshop YITP-T-19-03 "Quantum Information and String Theory 2019," where discussions about the early stages of this work were held. N.B. is supported by the National Science Foundation under grant number 82248-13067-44-PHPXH, by the Department of Energy under grant number DE-SC0019380, and by New York State Urban
Development Corporation Empire State Development contract no. AA289.

\end{document}